\documentclass[a4paper]{spie}  

\usepackage{amsmath,amsfonts,amssymb}
\usepackage[]{graphicx}

\usepackage{aas_macros}
\usepackage[colorlinks=true, allcolors=blue]{hyperref}

\title{MICADO PSF-Reconstruction work package description}

\author[a]{Matteo Simioni}
\author[a]{Carmelo Arcidiacono}
\author[a]{Andrea Grazian}
\author[b]{Yann Clenet}
\author[c]{Richard Davies}
\author[a]{Marco Gullieuszik}
\author[d]{Gijs Verdoes Kleijn}
\author[e]{Fernando Pedichini}
\author[f]{Roland Wagner}
\author[f,g]{Ronny Ramlau}
\author[h]{Werner W. Zeilinger}
\author[b]{Fabrice Vidal}
\author[a]{Benedetta Vulcani}
\author[a,i]{Roberto Ragazzoni}
\author[b]{Arnaud Sevin}
\author[a]{Bernardo Salasnich}
\author[a]{Andrea Baruffolo}
\author[j]{Lorenzo Busoni}
\author[j]{Simone Esposito}
\author[b]{\'{E}ric Gendron}
\author[e]{Roberto Piazzesi}
\author[k]{Elisa Portaluri}
\author[a]{Anita Zanella}
\author[l]{Tapio Helin}
\author[m]{Hanindyo Kuncarayakti}
\author[m]{Seppo Mattila}
\author[a]{Renato Falomo}
\author[j]{Enrico Pinna}
\affil[a]{INAF Osservatorio Astronomico di Padova, Vicolo dell'Osservatorio 5, I-35122, Padova, Italy}
\affil[b]{LESIA, Observatoire de Paris, Section de Meudon 5, place Jules Janssen, F-92195 Meudon Cedex, France}
\affil[c]{MPE - Max-Planck-Institut für extraterrestrische
Physik
Giessenbachstrasse 1,
D-85748 Garching,
Germany}
\affil[d]{University of Groningen, PO Box 72, 9700 AB Groningen, the Netherlands}
\affil[e]{INAF - Osservatorio Astronomico di Roma, Via Frascati 33, 00040, Monte Porzio Catone, Italy}
\affil[f]{Industrial Mathematics Institute, Johannes Kepler University Linz, Altenberger Strasse 69, 4040 Linz, Austria}
\affil[g]{RICAM - Johann Radon Institute for Computational and Applied Mathematics, Altenberger Strasse 69,
4040 Linz,
Austria}
\affil[h]{Department of Astrophysics, University of Vienna, Tuerkenschanzstrasse 17, A-1180, Wien, Austria}
\affil[i]{Dipartimento di Fisica e Astronomia, Universit\`a degli studi di Padova, via Marzolo 8, I-35131, Padova, Italy}
\affil[j]{INAF - Osservatorio Astrofisico di Arcetri, Via E. Fermi 5, 50125, Firenze, Italy}
\affil[k]{INAF - Osservatorio Astronomico d'Abruzzo, Via Mentore Maggini, I-64100 Teramo, Italy}
\affil[l]{	
LUT University,
P.O.Box 20,
FI-53851, Lappeenranta
Finland}
\affil[m]{Tuorla Observatory, Department of Physics and Astronomy, FI-20014 University of Turku, Finland}
\authorinfo{Further author information: (Send correspondence to M.S. and C.A.)\\M.S.: E-mail: matteo.simioni@inaf.it, Telephone: +39 049 829 3547\\  C.A.: E-mail: carmelo.arcidiacono@inaf.it, Telephone: +33 049 829 3414}

\begin{document} 
\maketitle

\begin{abstract}
The point spread function reconstruction (PSF-R) capability is a deliverable of the MICADO@ESO-ELT project. The PSF-R team works on the implementation of the instrument software devoted to reconstruct the point spread function (PSF), independently of the science data, using adaptive optics (AO) telemetry data, both for Single Conjugate (SCAO) and Multi-Conjugate Adaptive Optics (MCAO) mode of the MICADO camera and spectrograph. 
 The PSF-R application will provide reconstructed PSFs through an archive querying system to restore the telemetry data synchronous to each science frame that MICADO will generate. Eventually, the PSF-R software will produce the output according to user specifications.
The PSF-R service will support the state-of-the-art scientific analysis of the MICADO imaging and spectroscopic data.

\end{abstract}


\section{INTRODUCTION}
\label{sec:intro}  
MICADO\cite{micado2010} is the Multi-AO Imaging Camera for Deep Observations. It is one of the first light instruments of the ESO ELT\cite{elt2007,elt2008,E-ELT}, providing capability both for imaging and long-slit spectroscopy at near-infrared wavelengths while exploiting single conjugate adaptive optics\cite{Babcock253} (SCAO) and multiconjugate adaptive optics\cite{beckers88,beckers89a,LO1} (MCAO).
This provides a high level of flexibility for modes that are not offered by other ELT instruments, giving the opportunity to reach unique and major scientific results.

The laser guide star\cite{modaltomographyN,modaltomographyAA} MCAO system is currently being developed by the MAORY\cite{maory2010messenger} consortium, while the natural guide star (NGS) SCAO system is developed jointly by the MICADO and MAORY consortia\cite{clenet_implementation_2015}. MICADO will interface to the MAORY warm optical relay that re-images the telescope focus. In this configuration, both MCAO and SCAO are available. If required for an initial phase, MICADO will also be able to operate with just the SCAO system in a stand-alone mode, using a simpler optical relay that interfaces directly to the telescope. MICADO has four observing modes producing data: standard imaging, astrometric imaging, high-contrast imaging, and spectroscopy.
Each observing mode can optionally be operated in a configuration that gathers the data (wavefront sensor, WFS, images and telemetry) required for point spread function reconstruction (PSF-R). 

PSF-R is the process to determine the point spread function (PSF) of a science exposure, without using the focal plane science images. The reconstructed PSF is instead estimated using (i) images and telemetry generated by the AO system, (ii) information by other systems monitoring the atmosphere and (iii) telemetry from the ELT and MICADO instrument. 
The MICADO PSF-R is a service that complements, but is not part of, the MICADO data processing pipeline. It is a tool that assists data analysis rather than data reduction. But to date, no ground-based observatory offers PSF-R as a facility tool for AO observations. The development of the PSF-R is a collaborative effort involving involving the ELT and MAORY as well as MICADO, since all these systems contribute to the final measured PSF, and they all have a role in saving and archiving the necessary data.

The MICADO PSF-R service will support standard imaging, astrometric imaging, and spectroscopic observations. 
On the other hand, the PSF-R service is not planned for high-contrast imaging observations, as these focus on subtracting the PSF using ad-hoc techniques. The PSF-R will not support seeing enhanced and seeing limited (without AO) observations either. Finally, PSF-R is not suitable for non-sidereal imaging data in case a spatially-resolved moving target is used as AO guide star for the WFS.

The accuracy of the reconstructed PSF is driven by the scientific requirements of MICADO. The goals of the PSF-R service is to provide reconstructed PSFs with defined metrics (Strehl ratio, encircled energy, full width at half maximum, FWHM, ellipticity) within the reference values summarized in Table\,\ref{tab:refval}. The quoted accuracies have been derived by analytic calculations and they will be refined with simulations being carried out by the MICADO science team and by the scientific evaluation work package (WP) of the MICADO PSF-R team.

\begin{table}[ht]
  \caption{Error budget for the reconstructed PSF parameters. Error budget for PSF-R is indicated in the third and fourth column respectively for the nominal case and the operational case in extreme conditions (i.e. strong turbulence, faint NGS, science object far away from the science target, incomplete AO data).}
\label{tab:refval}
\begin{center}       
\begin{tabular}{|l|l|l|l|} 
\hline
\rule[-1ex]{0pt}{3.5ex}  parameter & notes & nominal accuracy & accuracy - extreme conditions  \\
\hline
\rule[-1ex]{0pt}{3.5ex} Strehl Ratio     & SCAO on-axis    & $\pm2\%$            & $\pm10\%$                         \\
\rule[-1ex]{0pt}{3.5ex}                  & SCAO off-axis   & $\pm10\%$           & $\pm20\%$ without $C_{n}^{2}$-profile \\
\rule[-1ex]{0pt}{3.5ex}                  & MCAO full field & $\pm2\%$            & $\pm10\%$                         \\
\hline
\rule[-1ex]{0pt}{3.5ex} Encircled Energy &                 & $\pm5\%$            & $\pm10\%$                         \\
\hline
\rule[-1ex]{0pt}{3.5ex} FWHM             &                 & $\pm5\%$            & $\pm10\%$                         \\
\hline
\rule[-1ex]{0pt}{3.5ex} Ellipticity      & PSF axis ratio at FWHM & $\pm5\%$ & $\pm10\%$ \\
\hline
\end{tabular}
\end{center}
\end{table}

\section{WORK PACKAGES AND TEAM MEMBERS}
\label{sec:wp}
The MICADO PSF-R team works on the implementation of the instrument software devoted to reconstruct the PSF, independently of the science data, using AO telemetry data, both for SCAO and MCAO mode of the MICADO camera and spectrograph. The work is distributed over $9$ WPs, listed in Table~\ref{tab:wpppl}, together with the institutions responsible for their development. In the following, the goals of each WP are summarized.

\begin{table}[ht]
\begin{center}
\caption{MICADO PSF-R WPs. For each WP, the responsible institution is also listed.}
\begin{tabular}{|c|l|l|l|}
\hline
\rule[-1ex]{0pt}{3.5ex} WP & WP name & Responsible institution & WP Reference \\
\hline
\rule[-1ex]{0pt}{3.5ex} 1 & PSF-R management and systems engineering & INAF-OAPD & \\
\hline
\rule[-1ex]{0pt}{3.5ex} 2 & PSF-R SCAO algorithm                     & JKU Linz  & \cite{2018JATIS...4d9003W} \\
\hline
\rule[-1ex]{0pt}{3.5ex} 3 & PSF-R MCAO algorithm                     & JKU Linz  & \cite{2018JATIS...4d9003W} \\
\hline
\rule[-1ex]{0pt}{3.5ex} 4 & SCAO simulations for PSF-R               & LESIA     & \cite{clenet_implementation_2015,clenet_micado_2018}               \\
\hline
\rule[-1ex]{0pt}{3.5ex} 5 & MCAO simulations for PSF-R               & INAF-OAPD & \cite{2018SPIE10703E..4IA,2020PASP..132h4502P}                     \\ 
\hline
\rule[-1ex]{0pt}{3.5ex} 6 & Non-AO PSF-R algorithms and calibrations & INAF-OAR  & \\ 
\hline
\rule[-1ex]{0pt}{3.5ex} 7 & PSF-R data flow architecture             & NOVA      & \cite{2015scop.confE..51V,2020arXiv201104516B}                     \\ 
\hline
\rule[-1ex]{0pt}{3.5ex} 8 & PSF-R user software                      & INAF-OAR  & \\ 
\hline
\rule[-1ex]{0pt}{3.5ex} 9 & PSF-R scientific evaluation              & INAF-OAPD & \cite{2014A&A...568A..89G,2016A&A...593A..24G}                     \\ 
\hline
\end{tabular}
\end{center}
\label{tab:wpppl}
\end{table}

The WP1 (PSF-R management and systems engineering) is responsible for the communication within the PSF-R team and coordinating the project. It is in charge of the managerial and sociological aspects within the team, of the development plan and its coordination with the rest of the MICADO and MAORY schedule. 

The algorithms for the reconstruction of the SCAO PSF are developed within WP2 (PSF-R SCAO algorithm). The data items, that need to be provided by the Real-Time Control (RTC), and the relevant calibrations are defined, in coordination with the MICADO SCAO team. Test cases provided by the SCAO simulation WP (WP4) are used to check the performance of the SCAO algorithms. Splitting simulations and reconstruction tasks avoids development biases. The error budget is estimated by measuring the noise of the input data and propagating the errors throughout the algorithms. This WP contributes the SCAO reconstruction module to the final PSF-R service.

Similarly, WP3 (PSF-R MCAO algorithm) provides the algorithms to reconstruct the PSF for the MCAO configuration with MAORY. The development will be done in close collaboration with the MAORY team. It should be noticed that MAORY follows a different schedule than MICADO, which might have an impact on the work planning.

As already mentioned, the simulations of the AO systems should follow as close as possible the development of the systems themselves. Therefore, WP4 (SCAO simulations for PSF-R) is split from the reconstruction algorithms and assigned to the team building the AO system. In this way the in-depth knowledge of the instrument development team can be used and increases the confidence that the PSF reconstruction will perform as expected on the final system. Additionally, the development of the SCAO system is supported by a dedicated optical bench breadboard and an on-sky test. Both these experiments can improve the insight on the PSF reconstruction performance and the technical readiness of the algorithms. So it is crucial for the PSF-R development to be part of those tests. The WP4 is waiting for test requests from the SCAO algorithm development team.

For WP5 (MCAO simulations for PSF-R) similar considerations apply as in the case of WP4, so it will be assigned to the MAORY team. There is no dedicated bench experiment foreseen in the MAORY development at the moment, but there is the possibility for a test during the integration phase before shipping MAORY to Chile (planned for 2025). Also WP5 is waiting for test requests from the MCAO algorithm development team.

There are several systems influencing the PSF beside the classically considered AO system. The WP6 (Non-AO PSF-R algorithms and calibration) is concerned with these other impacts on the PSF and investigates their influence, the data available and the calibrations necessary for compensation. With an error budget all the terms are balanced and the work on the specific elements prioritised. The output algorithms of this work package are either used as input for the AO PSF algorithms or to post-process the PSF. The necessary data items for the algorithms will be defined and communicated to the WP7 for proper collection.

The WP7 (PSF-R data flow architecture) works on the design of the data flow system that starts from the raw data in the various systems, as defined by the algorithmic work packages, and ends with the storage in the PSF-R data archive. This work and the subsequent qualification of the implementation is performed in close collaboration with ESO.

The WP8 (PSF-R user software) will integrate the PSF-R modules and algorithms developed in the algorithmic work packages in a user software. The software has to be written and documented as agreed by ESO so that it can be accepted for takeover.

Finally, the goal of the WP9 (PSF-R scientific evaluation) is to provide the algorithmic development (the WP2 and WP3)  with performance metrics, define science test cases and evaluate the impact of the PSF reconstruction algorithms on the science test cases.


\section{PSF RECONSTRUCTION}
\label{sec:PSFR}
\subsection{Why the PSF-R is needed}
Given the unprecedented size of the ELT, the resulting PSF is quantitatively different from that of any present AO facility. In fact, the core of the ELT PSF is a factor $\sim 5$ narrower in diameter. This means that with respect to the core, and at the same Strehl ratio, (i) the surface brightness of the halo is a factor $\sim 5^{2}$ fainter, and (ii) the size of the halo is a factor $\sim 5$ larger. This means that, for most stars the halo is barely detected as a distinct feature, and that multiple halos blend into a faint distributed background. On the other hand, the first Airy ring will be a prominent feature of the PSF halo.

These conditions make the measure of the total flux of single sources particularly challenging, without external input (like e.g. an indication of the Strehl ratio), consequently affecting photometric measurements. The lower is the achieved Strehl ratio of the observation, the higher is the contribution of the PSF halo to the total flux of the source.

The PSF-R philosophy adopted for MICADO is to reconstruct a PSF independently of the science data, to have the widest applicability (see Wagner et al. 2019\cite{wagner:hal-02624937} for a review on PSF determination techniques). 
In fact, it has been estimated, referring to extragalactic targets, that in the $35$-$40\%$ of cases, a MICADO science frame is void of point sources suitable for standard PSF characterization\cite{ric_psfrcontext} .
In MICADO PSF-R, the AO telemetry is used to model the optical transfer function (OTF), together with nearby $C_{n}^{2}$ profiler data, and on-sky phase diversity. Specifically, an initial PSF estimate is  adapted according to various degradation effects and other calibrations, to provide a final reconstructed PSF. It is important to handle separately the AO and non-AO effects, as their scaling with wavelength is different. The residual wavefront leads to a wavelength dependence of the PSF shape; residual chromatism affects the location of the PSF at each wavelength; and residual vibration is independent of wavelength. The plan for MICADO is that these effects are calibrated or measured so that they can be included appropriately in the final reconstructed PSF. Crucially, the data flow architecture has to be compatible with ESO’s data-flow and archiving systems, especially on how the various data are gathered, saved, and made available to the user. In addition, a significant effort is being made to derive non-AO contributions in a way that minimizes on-sky calibrations.

\subsection{Algorithm for PSF-R}
The PSF obtained by the optical complex system made of the chain of the turbulent atmosphere, the ELT telescope system, the MICADO relay optics, the SCAO system and the MICADO science instrument will not be perfect. Along with the instrument specifications, the SCAO system will deliver a PSF whose Strehl ratio is up to $60\%$ (Goal: $70\%$) in K-band in median atmospheric turbulence conditions with a bright NGS in the close vicinity of the science target. The optical performance will be reduced in less optimal cases (strong turbulence, less bright NGS, science object far away from the science target).

PSF-Reconstruction algorithms mostly operate to produce an OTF (i.e. the Fourier space counterpart of the PSF). The OTF of the PSF part related to atmospheric turbulence in a specific direction and wavelength is computed to reconstruct the post-AO PSF associated with a given MICADO SCAO scientific observation. The OTF can be calculated from the so-called structure function ($D$) which is related to the residual incoming phase. The approach from Veran et al. (1997)\cite{1997JOSAA..14.3057V} is followed, splitting the structure function into several parts which are assumed to be independent. Each of these parts will be wavelength dependent and use calibration data items. The final result is the combined post-AO OTF for a specific wavelength $\lambda$ and direction $\alpha$ associated to one observational frame. It needs to be multiplied by the non-AO OTF and Fourier transformed to obtain the PSF. All computations are performed on the same resolution otherwise interpolations will add ghosts when applying a Fourier Transform. As regular grids for sampling the incoming phase are used, particular care is given in making sure that with a hexagonal actuator pattern, all frequencies which can be represented on M4 are still sampled. Using the currently available ESO data package, this need is fulfilled with a grid spacing of $0.25$~m projected on M1. Furthermore, the phase $\phi$ is related to the wavefront $\varphi$ via $\phi =\frac{2\pi}{\lambda}\varphi$ and thus wavelength dependent.

The algorithm developed in the context of the MICADO PSF-R is currently composed of the following blocks:
\begin{enumerate}
\item Wavefront low order structure function. From the reconstructed residual incoming phase $\phi_{rec,t}$, for time frames $t$ and related to the direction of the NGS at wavelength $\lambda$, the 2D averaged structure function $D_{rec}$ is computed. $\phi_{rec,t}$ is computed from the WFS measurements using the AO control algorithm. Note that this computation can be done as a pre-processing step. Alternatively the computed deformable mirrors (DM) shapes (i.e. M4 + M5 commands in SCAO) can be used to extract this information.
\item Aliasing structure function. A model is used to propagate simulated higher order parts of phases through the WFS and then compute the response of the control algorithm. The output is the structure function of the aliasing part of the phase $D_{al}$. This procedure is rerun for multiple time steps before all these computed aliasing phases are used in a structure function computation. The input list for this algorithm includes seeing, $r_0$, used AO control matrix/algorithm and Pupil mask $P$.
\item Noise structure function. The AO control algorithm is used to propagate the noise model from the WFS level to wavefront level $D_{n}$. This computation needs the following inputs: guide star magnitude (or photon flux level) at $\lambda$; Pupil mask $P$ (from calibrations); M4 actuator positions; deformable mirrors influence functions and active WFS subapertures.
\item Wavefront higher order structure function. From a statistical model of the atmospheric turbulence (von Karman, Kolmogorov) a structure function for the higher order part not corrected by the AO system, $D_{\perp}$, is computed. This cannot be done once for all since the pupil mask changes. 
\item Anisoplanatic structure function. Using the reconstructed residual incoming phases $\phi_{t,rec}$, and knowledge on the atmospheric turbulence profile, an atmospheric tomography can be performed to recover the atmospheric turbulence residual in the direction of interest. The difference between this residual and the reconstructed residual wavefronts during the AO run, the anisoplanatic structure function $D_{an}$ can be computed.
\item Combined post-AO OTF. From each of the above structure functions ($D_{rec}$, $D_{al}$, $D_{n}$, $D_{\perp}$ and $D_{an}$) an OTF has to be computed as 
$$OTF_x = \exp (-\frac{1}{2} D_x)$$
with $x \in \{\|, \perp, an\}$, where $D_\| = D_{rec} - D_{n}+D_{al}$. The telescope OTF ($OTF_{tel}$), instead, is computed from $P$ and these OTFs are combined by multiplication to obtain the full AO OTF: $$OTF_{AO} = OTF_{tel} \cdot OTF_{\|} \cdot OTF_\perp\cdot OTF_{an}$$.
\end{enumerate}

Finally, the Non-AO effects are also added. They are composed by two parts: the non-common path aberrations (NCPA) and the vibration pattern. 
The NCPA are a series of Zernike coefficients ($Z_{3},...,Z_{100}$) that are transformed into an OTF function defined as $OTF_{NCPA}$: they model the NCPA for that specific instrument configuration and telescope pointing.
The vibration pattern is a 2D array where each element is the integral over the exposure time of the presence of the chief ray in that element, and the whole array is normalised to $total(array)=1$: in other words, it is equivalent to a frame in which the value of each pixel is proportional to the number of times the chief ray has passed over it.
The final PSF, $PSF_{MICADO}$, is retrieved in three steps for each point of the field:
\begin{enumerate}
    \item Compute $OTF_{NCPA}$ from Zernike coefficients.
    \item The $OTF_{AO}$ is combined with the $OTF_{NCPA}$ to calculate the $OTF_{AO+NCPA}$ as $$OTF_{AO+NCPA}=OTF_{AO}\cdot OTF_{NCPA}$$
    \item The $OTF_{AO+NCPA}$ is transformed via $FFT^{-1}$ into the $PSF_{AO+NCPA}$
    \item The kernel of the residual vibrations pattern $K_{vibration}$ is registered to the focal plane coordinates, and convolved with the $PSF_{AO+NCPA}$ to obtain the final $PSF_{MICADO}$ as $$PSF_{MICADO}=PSF_{AO+NCPA} * K_{vibration}$$
\end{enumerate}

\section{EXAMPLE OF PSF RECONSTRUCTION}
\label{sec:LBT}
\begin{figure}
  \centering
  \includegraphics[width=0.9\linewidth]{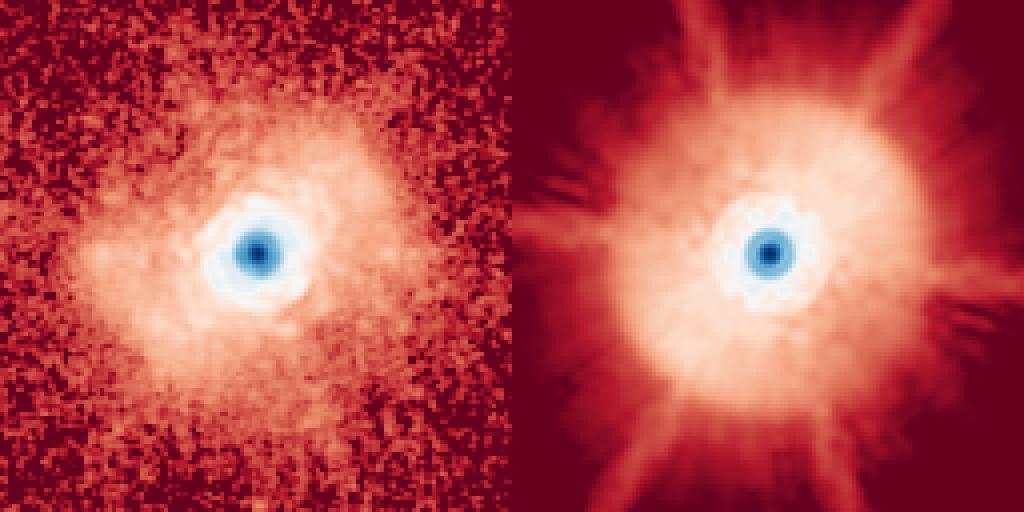} 
      \caption{The picture shows: (Left) the LUCI observed PSF, (Right) the reconstructed PSF. The frame angular size is $3".8\times 3".8$.}
         \label{fig:lbt}
  \end{figure}

\begin{figure}
  \centering
  \includegraphics[width=0.49\linewidth]{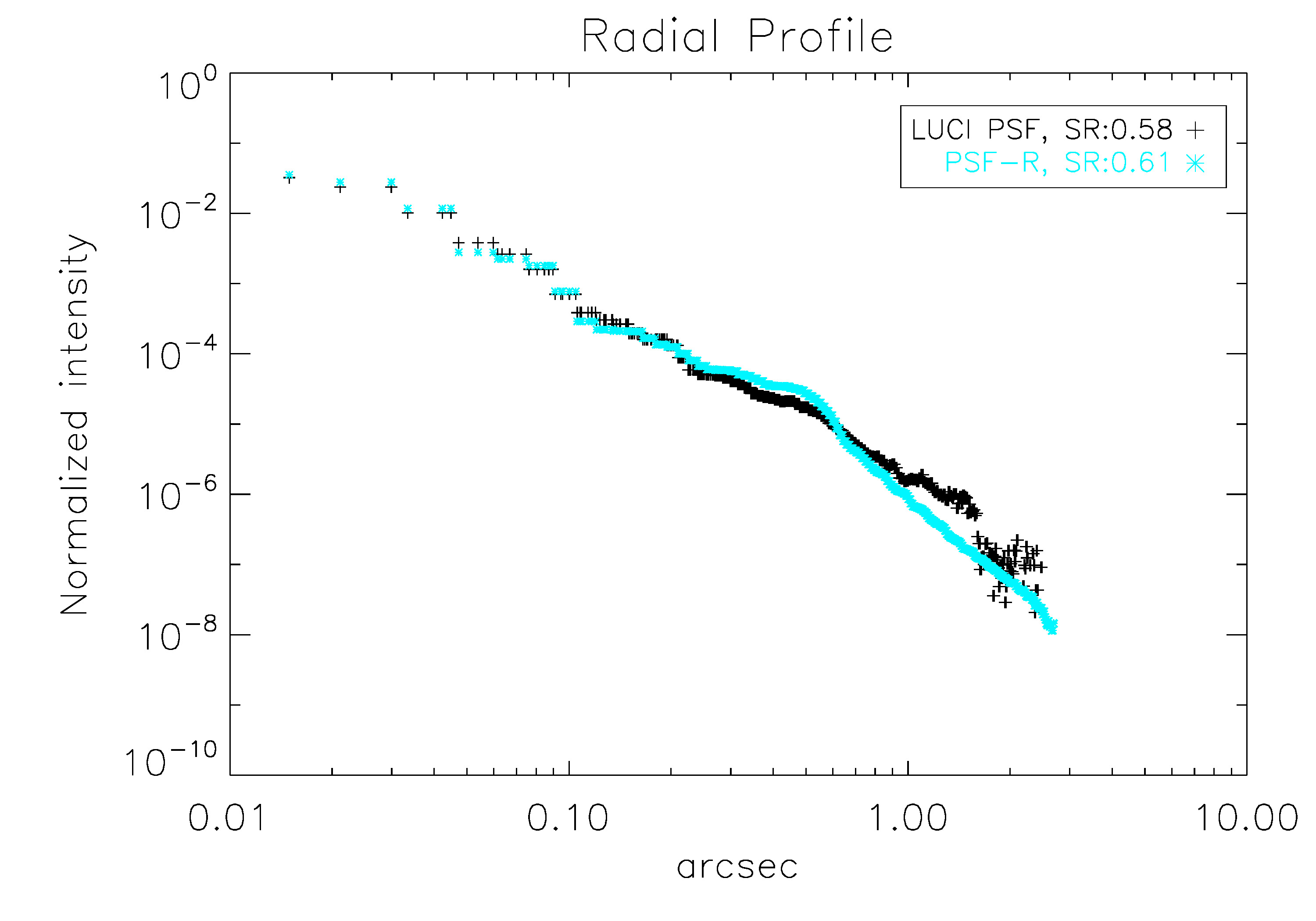}
  \includegraphics[width=0.49\linewidth]{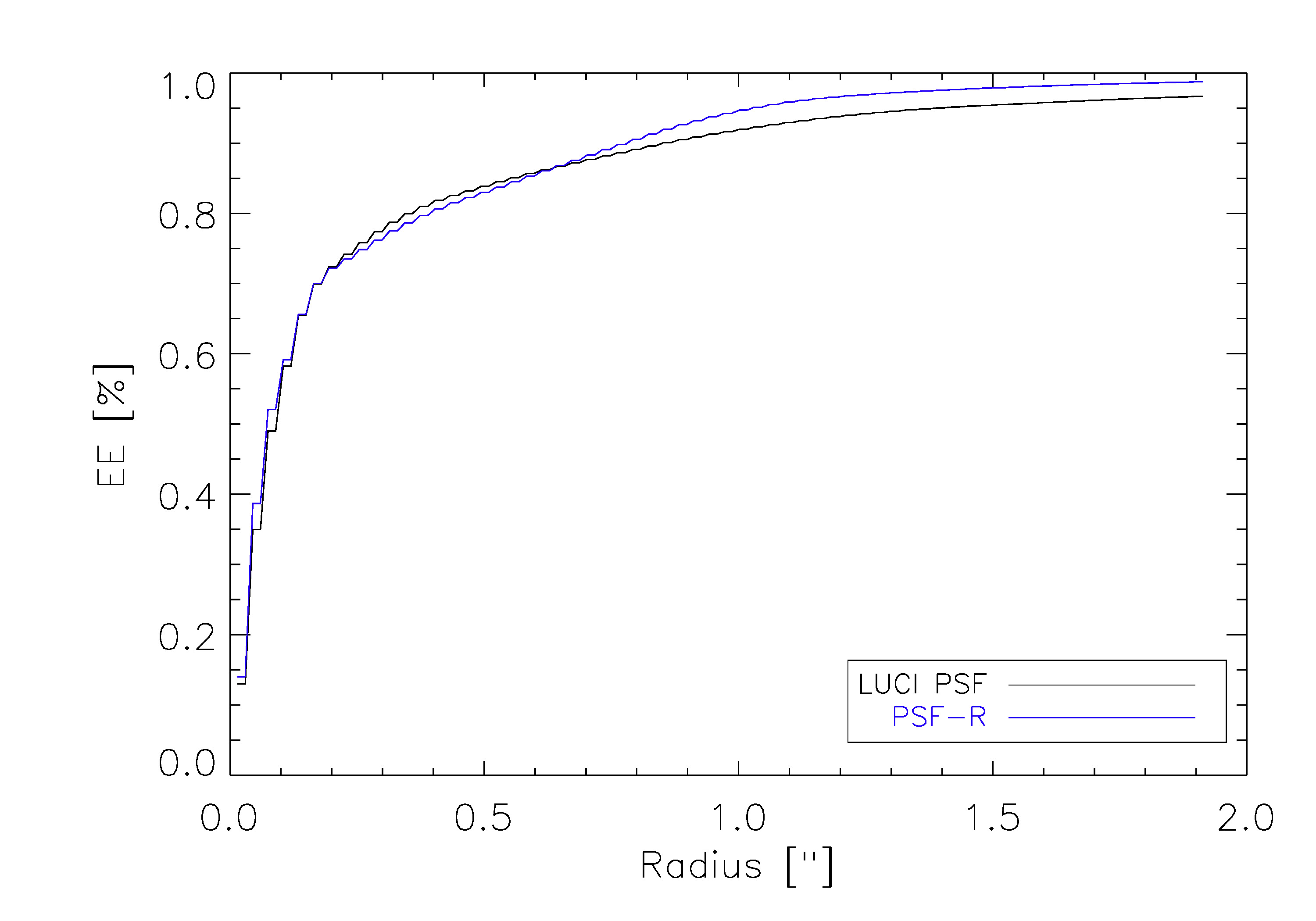} 
      \caption{Intensity and encircled energy profiles.}
         \label{fig:profile}
  \end{figure}

The first prototype of the PSF-R algorithm has been tested using closed loop telemetry and near infrared images acquired with the SOUL\cite{pinna_soul_2016} system mounted on the front bent Gregorian focus of the LBT\cite{2010ApOpt..49..115H} telescope and the LUCI\cite{seifert_lucifer_2003} camera. A diffraction limited camera (a.k.a. N30) equips the LUCI instrument for both imaging and spectroscopy, offering a $2k\times 2k$ sensitive area corresponding to $30"\times 30"$ on the sky (14.95 mas/px). For this preliminary testing of the PSF-R algorithm, laboratory-like condition has been used: optical turbulence disturbance is generated and corrected\cite{FLAO} using the Adaptive Secondary AdSec\cite{riccardi2003} of the LBT, with the telescope and AO system configured for daytime\cite{acc_test}. The SOUL team provided the essential data set for the PSF-R: AO WFS slopes data history, control matrix, interaction matrix, gain vector, and the pupil definition. The PSF-R does not use the camera observing data for the analysis. The LUCI observed PSF was only used as comparison for the analysis of the performed PSF-R. Figure\,\ref{fig:lbt} shows the matching of the observed and reconstructed PSF. The central bright area corresponds to the spatial frequencies controlled by the AdSec and limited by inter-actuator distance. Beyond, the uncorrected dome seeing turbulence contributes are negligible.  More quantitatively, in Figure\,\ref{fig:profile} it can be noticed the very good correspondence of the measured and reconstructed Strehl ratio (FeII filter) of 0.58 and 0.61 respectively. The PSF-R algorithm also reconstructed the FWHM and encircled energy profiles within the measurements errors (within a few percent points). 
  
\section{SUMMARY}
\label{sec:sum}
The PSF-R capability is a deliverable of the MICADO@ESO-ELT project. This will make MICADO the first instrument to date to offer this service. In order for the deliverable to have the widest scientific applicability, a pure PSF-R approach is pursued, i.e. to reconstruct the PSF independently of the science data.
The MICADO PSF-R team is working on the development of the PSF-R method with $9$ WPs actively interacting. Their aims and responsibilites have been presented, along with a description of the PSF-R algorithm.
The main features of the MICADO PSF-R service include:

\begin{itemize}
\item[--] Generate field-dependent reconstructed PSF at the desired wavelength, taking into account both AO and non-AO (e.g. telescope or atmospheric) input.
\item[--] Provide reconstructed PSFs through an archive querying system to restore the telemetry data synchronous to each of the science frames that MICADO will generate.
\item[--] Produce the output according to user specifications.
\item[--] Support the state-of-the-art scientific analysis of the MICADO imaging and spectroscopic data (both for SCAO and MCAO observing mode, the latter provided by the MAORY module).
\end{itemize}

Finally, results coming from the application of the first prototype of the PSF-R algorithms on real SOUL@LBT data have been presented. The obtained reconstructed PSF shows small deviation from the observed one, of the order of few percent points in Strehl ratio, FWHM and encircle energy profile. Remaining within the measurement errors, these results are encouraging already at the present stage and they are a good starting point toward a successful development of the PSF-R algorithm for MICADO. 
The PSF-R team of MICADO will work in the next $\sim5$ years to further improve these algorithms. The general philosophy of MICADO PSF-R could be exported also to other AO facilities of ESO-ELT and other gigantic telescopes in the future.

\acknowledgments 
This work has been partly supported by INAF through the Math, ASTronomy and Research (MAST\&R), a working group for mathematical methods for high-resolution imaging.
The PSF-R Team thanks Miska Le Louarn (ESO) for continued support and enthusiasm for the project.

\bibliography{report} 
\bibliographystyle{spiebib} 

\end{document}